\documentclass[onecolumn,a4paper,12pt]{article}

\usepackage{amssymb}
\usepackage{amsmath}
\usepackage{graphicx}
\usepackage[a4paper, landscape, margin=1in]{geometry}
\usepackage{longtable}

\begin{document}

\title{Lempel-Ziv complexity analysis of one dimensional cellular automata}

\author{E. Estevez-Rams\\
Facultad de F\'isica-IMRE,\\
University of Havana, San Lazaro y L. CP 10400.
La Habana. Cuba.\\
Corresponding author. e-mail: estevez@imre.oc.uh.cu\\
\and
R. Lora-Serrano\\
Universidade Federal de Uberlandia, AV. \\
Joao Naves de Avila, 2121- Campus Santa Monica, CEP 38408-144,\\
Minas Gerais, Brasil.\\
\and
C. A. J. Nunes\\
Universidade Federal de Uberlandia, AV. \\
Joao Naves de Avila, 2121- Campus Santa Monica, CEP 38408-144,\\
Minas Gerais, Brasil.\\
\and
B. Arag\'on Fern\'andez\\
Universidad de las Ciencias Inform\'aticas (UCI),\\
Carretera a San Antonio, Boyeros. La Habana. Cuba.
}

\date{\today}
\maketitle

\begin{abstract}
Lempel-Ziv complexity measure has been used to estimate the entropy density of a string. It is defined as the number of factors in a production factorization of a string. In this contribution we show that its use can be extended, by using the normalized information distance, to study the spatiotemporal evolution of random initial configurations under cellular automata rules. In particular the transfer information from time consecutive configurations is studied, as well as the sensitivity to perturbed initial conditions. The behavior of the cellular automata rules can be grouped in different classes but no single grouping captures the whole nature of the involved rules. The analysis carried out is particularly appropriate for studying the computational processing capabilities of cellular automata rules. 
\end{abstract}

\begin{quotation}
Cellular automata (CA) have long attracted attention as dynamical systems with local updating rules and yet can exhibit, for certain rules, complex, long space and time correlated patterns. This contrast with other rules which results in trivial patterns being homogeneous or periodic. In this article we approach CA from two related angles: we analyze the information transfer in the time evolution of CA driven sequences and; we revisit the sensibility of the initial configuration on sequence evolution. In order to do so, we borrow a recently reported information distance based on Kolmogorov algorithmic complexity. The normalized information distance has been used previously to find a hierarchical clustering of CA rules. What is different in our approach, is the temporal analysis of the sequence evolutions by correlating different calculated distances with entropy density.  Entropy rate, is a length invariant measure of the amount of irreducible randomness in a dynamical process.  In order to perform our analysis, we incorporate to the practical calculation of the entropy rate and the distance measure, the use of Lempel-Ziv complexity. Lempel-Ziv complexity carries a number of practical advantages while avoiding the uncomputable nature of Kolmogorov randomness. The reduction of entropy density during time evolution can be related to energy dissipation through Landauer principle. Related to the last fact, is the computational capabilities of  CA as information processing rules, were the performed analysis could be used to select CA rules amiable for simulating different physical process. The tools developed in this article for the analysis of the CA are easily extendible to the study of other one dimensional dynamical systems.
\end{quotation}

\section{Introduction}

Emergence of complex patterns out of, by all accounts, trivial initial conditions in large systems has been the subject of intensive studies (See \cite{crutchfield12} and reference therein). The question can arise when modeling a broad range of real systems such as those found in physics, chemistry or biology. Complex patterns are considered far from trivial periodic behavior (including the period 1 or fully homogeneous system), as well as the completely random system. Periodicity has a low information content, contrary to randomness, yet both are statistically simple to describe \cite{crutchfield92}. Another way of looking at the idea of complexity is the existence of large spatiotemporal correlations in the system dynamics \cite{wolfram86}.

From a mathematical point of view, discrete time and space cellular automata (CA) has driven much attention over the past decades ( See \cite{kari05} and reference therein). CA can show from trivial periodic behavior to universal computing capabilities \cite{wolfram84}. Given a finite number of states $p$, and a finite interaction range $r$, CA evolves according to local rules and yet, can achieve large spatiotemporal correlations. When $p=2$ and $r=1$ \footnote{A $r$ range interaction implies a $2r+1$ neighborhood} CA are known as elementary CA or ECA, defined over $2^{2^{3}}=(256)$ possible local rules \cite{wolfram86}. 

Crutchfield and Shalizi have used the machinery of computational mechanics to explore local and global characteristics of CA dynamics \cite{shalizi01,hordijk01}. Computational mechanics aims at discovering the emergence of structure, and is specially well suited to explore the computational capabilities of a dynamical system \cite{crutchfield12}.

Others have used Kolmogorov or algorithmic complexity  to analyze CA \cite{dubaq01}. Kolmogorov complexity characterize a system by the length of the  shortest algorithm running on a Universal Turing Machine, capable of reproducing the system \cite{kolmogorov65,vitanyi93}. It is an absolute measure, up to a constant value, and exhibits useful properties as a measure of randomness ( in what follows we will use the term Kolmogorov randomness). The main drawback of Kolmogorov randomness is its uncomputability as a result of the halting problem \cite{vitanyi93}. This lead to the need of defining practical alternatives based on the compressibility of the mathematical description of the system. In the analysis of CA by Kolmogorov randomness, the system is left to evolve for an (arbitrary) long time and the final configurations are characterized by its compressibility \cite{dubaq01}. 

CA can exhibit sensibility to initial conditions. For two different initial conditions, how does the CA evolves to increasingly (or not) different configurations can be characterized, for example, defining 'Lyapunov' coefficients \cite{wolfram84,shreshevsky92,Bagnoli199234}. Another approach is to analyze the evolution of the compressibility property of the two systems \cite{zenil10}. 

Emmert-Streib has used a distance function based on Kolmogorov randomness and introduced in \cite{li04}, to derive a hierarchical clustering of ECA \cite{emmert10}. Due to the fact that Kolmogorov randomness is uncomputable, the use of some compression software such as gzip or bzip2 is taken as a practical alternative. The use of compression software to estimate Kolmogorov randomness can be criticized by a number of reasons, one being the necessarily finite size of the words dictionary \cite{weinberger92}, which leads to limitations in the size of the systems analyzed depending precisely in the Kolmogorov randomness of the sequence, the very quantity aimed to be estimated. In his approach, Emmert-Streib considered small configurations of CA with 50 cells and $10^3$ time steps. The system spatiotemporal evolution is  analyzed as a unique string by concatenating sequences derived from the temporal evolution and compression is performed over such mapping \cite{emmert10}.

In this paper we study the dynamics of CA by looking at the complexity of its spatial configurations using Lempel-Ziv complexity based quantities. Lempel-Ziv complexity  \cite{lz76} (from now on LZ76 complexity) is defined over a factorization of a string based on a production rule. LZ76 complexity measure has been used to analyze data sequences from different sources, theoretical or experimental \cite{szczepanski04,abo06,zhang09,contantinescu06,chelani11,rajkovic03,liu12,talebinejad11}. An important result that Ziv showed, is that the asymptotic value of the LZ76 complexity growth rate (LZ76 complexity normalized by $n/\log{n}$, where $n$ is the length of the sequence) is related to the entropy rate (as defined by Shannon information theory) for an ergodic source \cite{ziv78}. This theorem is essential at using the LZ76 complexity as an entropy rate estimator for finite sequences. Entropy rate has a close relationship with Kolmogorov randomness \cite{calude02}, and measures the irreducible randomness of a system \cite{feldman08}. We further extend the analysis to borrow the distance measure by Li et al \cite{li04}, and used by Emmert-Streib \cite{emmert10}, but using  LZ76 as the complexity measure. The approach avoids the limitations in system size, inherent in using compression software and allows to analyze large systems. In our analysis, we use $10^4$ cells evolving over the same number of time steps. We thus avoid any concatenation of configurations, as we are now capable of analyzing each temporal configuration with statistical meaning and therefore follow the time evolution of the CA.

The paper is organized as follows, in sections \ref{sec:ECA}, \ref{sec:lz76} and \ref{sec:NCD} the theoretical framework is laid, fixing notation and introducing the magnitudes used in the analysis. Section \ref{sec:results} reports the results obtained from the computer simulations. Both, the analysis of the evolution of an initial configuration under a CA  rule, as well as the sensitivity to initial conditions are discussed. This is followed in section \ref{sec:discuss} by discussion and conclusion. An appendix section gives useful tables.

\section{Elementary cellular automata}\label{sec:ECA}

A spatial and temporal discrete, one dimensional cellular automata, can be formally defined by a tern $(\Sigma, s, \Phi)$, where $\Sigma$ is a finite alphabet  of cardinality $\sigma(=|\Sigma|)$; $s=s_{0},s_{1},\ldots s_{N-1}$ is a set of sites and; $\Phi$ is a local updating rule. $s(i,j)$ denotes the substring $s_i s_{i+1}\dots s_j$ taken from $s\;(s(i,j) \subset s)$ and of length j-i+1. It is understood that if $j>N$ we take up to the last character $s_N$ of $s$. If $i>j$ then $s(i,j)$ will be the empty string. 

If $s^{t}=s^{t}_{0},s^{t}_{1},\ldots s^{t}_{N-1}$ denotes a particular configuration of the sites values at time $t$, then
\begin{equation}
 s^{t}_{i}=\Phi\left [ s^{t-1}(i-r, i+r)\right ].
\end{equation}
For elementary CA, $r=1$, $\sigma=2$ and a binary alphabet $\Sigma=\{0,1\}$ is used. 

There are a total of $2^{2^3}=256$ possible rules for ECA which can be labeled by a number. To each rule $\Phi$ a label $R$ is assigned according to a scheme proposed by Wolfram that has become standard \cite{wolfram02}:
\begin{equation}
 R=\Phi(0,0,0)2^{0}+\Phi(0,0,1)2^{1}+\Phi(0,1,0)2^{2}+\ldots +\Phi(1,1,1)2^{7}.
\end{equation}

ECA rules can be partitioned into equivalence class as a result of mirror and reversion symmetries\cite{wolfram02}, analysis can then be reduced to a representative member of each class. Equivalence class are shown in table \ref{tbl:WECAequiv} in the Appendix.

CA have been classified in a number of ways \cite{kari05}. The most cited one is the original classification of Wolfram \cite{wolfram84}. Starting from a arbitrary random initial configuration, CA are classified as: 
\begin{itemize} 
 \item [W1]: configurations evolves to an homogeneous state;
 \item [W2]: configurations evolves to a periodic behavior; 
 \item [W3]: configurations evolves to aperiodic chaotic patterns;
 \item [W4]: configurations evolves to configurations with complex patterns and long lived, correlated localized structures. 
\end{itemize}
Wolfram classification is vague and, furthermore, has been proven to be undecidable \cite{culik88}. As a result, the assignment of each rule to a Wolfram class is ambiguous. The classification is further complicated by the fact that some rules show high sensitivity to initial conditions, and different initial configuration can lead to different behaviors \cite{wolfram02}. We will not be particularly interested in classifying CA, yet for clarity, when referring to Wolfram classification  the table \ref{tbl:WECAclaf} found in the appendix will be used, which has been compiled from Wolfram Alpha knowledge engine \cite{wolframalpha}.

\section{Lempel-Ziv factorization and complexity}\label{sec:lz76}

Let the ''drop'' operator $\pi$ be defined as 
\begin{equation} 
s(i,j) \pi=s(i,j-1) \nonumber
\end{equation}
and, consequently,
\begin{equation}
s(i,j) \pi^k=s(i,j-k). \nonumber
\end{equation}

The Lempel-Ziv factorization \footnote{There are different Lempel-Ziv scheme for factorization \cite{cover06} and the reader must be careful to recognize which scheme is used in each case.} $F(s)$ of the string $s$ of length $N$
\begin{equation}
F(s)=s(1,l_1)s(l_1+1,l_{2})\dots s(l_{m-1}+1,N), \nonumber
\end{equation}
in $m$ factors is such that each factor $s(l_{k-1}+1,l_k)$ complies with
\begin{enumerate}
 \item $s(l_{k-1}+1,l_k)\pi\subset s(1,l_k)\pi^2$
 \item $s(l_{k-1}+1,l_k)\not\subset s(1,l_k)\pi$ except, perhaps, for the last factor $s(l_{m-1}+1,N)$.
\end{enumerate}
The first condition defines $F(s)$ as a history of $s$, while the second condition defines such history as an exhaustive history of $s$. The partition $F(s)$ is unique for every string \cite{lz76}.

For example the exhaustive history of the sequence $u=11011101000011$ is $F(s)=1.10.111.010.0001.1$,
where each factor is delimited by a dot.

The LZ76 complexity $C_{LZ}(s)$ $(=|F(s)|)$ of the sequence $s$, is defined as the number of factors in its exhaustive history. In the example above, $C_{LZ}(s)$=6.

In the limit of very large string length , $C(s)$ is bounded by \cite{lz76}
\begin{equation}\label{asymp}
 C_{LZ}(s) < \frac{N}{\log{N}}.
\end{equation}

Let the entropy rate be given by 
\begin{equation}\label{shannonh}
 h(s)= \lim_{N\rightarrow\infty}\frac{H[s(1,N)]}{N},
\end{equation}
where $H[s(1,N)]$ is the Shannon block entropy \cite{cover06} of the string $s(1,N)\in s$. Then, defining
\begin{equation}\label{lzh}
 c_{LZ}(s)=\frac{C_{LZ}(s)}{N/\log{N}}.
\end{equation}
Ziv \cite{ziv78} proved that, if $s$ is the output from an ergodic source, then
\begin{equation}\label{zivtheorem}
 \limsup_{N\rightarrow\infty}c_{LZ}(s)=h(s).
\end{equation}

This is the base of using $c_{LZ}$ as an estimate of $h_{\mu}$ for $N\gg 1$. 

Equation (\ref{zivtheorem}) is valid in the infinite limit, in practical cases equation (\ref{lzh}) is used as an estimate for the entropy density. Assuming a i.i.d. Gaussian distribution of the word length, the error estimate for the entropy density estimated by Lempel Ziv can be calculated from \cite{amigo06}
\begin{equation}
 \sigma=c_{LZ}^{3/2} \frac{d}{\sqrt{N \log N}}\label{eq:err}
\end{equation}
$d$ is the standard deviation of the word length distribution. For a $10^4$ length sequence, which will be the ones used in this study, equation (\ref{eq:err}) gives an order of magnitude for the error bound around $10^{-2}$. Lesne et al \cite{lesne09} has further discussed the estimation of error in the use of entropy estimators in short sequences. Using numerical simulations they showed that Lempel-Ziv estimators of entropy density show good agreement even for sequence as short as $2\times 10^3$. As already stated, in the simulations that follow, sequence length will be at least $10^4$ in length, and errors in the estimation of entropy density should then be of no significance.  

\section{Kolmogorov based normalized information distance}\label{sec:NCD}

The Kolmogorov randomness $K(s)$ of a string $s$, is the length of the shortest program $s^{*}$ that when run in a Universal Turing Machine (UTM), gives as output the string,
\begin{equation}
 K(s)=|s^{*}|.
\end{equation}
The UTM is stipulated to make the definition an absolute measure, up to a constant factor. Kolmogorov randomness captures the idea that a constant sequence can be described by a fairly small program, while a long enough sequence given by the toss of coin, can only be reproduced by the sequence itself and therefore will have large (as large as the sequence length itself) Kolmogorov randomness. Correspondingly, the conditional Kolmogorov randomness $K(s|p)$ is the length of the shortest program that can compute $s$ from the string $p$. Finally, the joint Kolmogorov randomness $K(s,p)$ is the size of the smallest program that computes both strings $s$ and $p$. There are several 'flavors' for the type of programs allowed in the Kolmogorov randomness definition, this does not make much difference in what follows. For the theoretical definitions, the reader can consider the prefix-free variant, where no program is a proper prefix of another program \cite{vitanyi93}. Kolmogorov randomness is uncomputable, a fact related to the halting problem, this is the biggest limitation to its practical use. It holds that
\begin{equation}
 K(s,p)\cong K(s)+K(p|s*)=K(p)+K(s|p*),\label{eq:ksp}
\end{equation}
where $\cong$ denotes that equality is valid up to a constant value independent of $p$ and $s$. 

Entropy density can be estimated from
\begin{equation}
 h(s)=\lim_{|s|\rightarrow \infty}\frac{K(s)}{|s|}.
\end{equation}
Even if the Kolmogorov randomness is uncomputable, the entropy density can be computed as can be seen from the results in the previous section.

The information about $s$ contained in $p$ is defined by
\begin{equation}
 I(s:p)=K(s)-K(s|p^{*}).
\end{equation}
Equation (\ref{eq:ksp}) implies that $I(s:p)=I(p:s)$ up to a constant. 

Given two sequences $s$ and $p$, Li et al. \cite{li04} defined the normalized information distance (NID) by the relation:
\begin{equation}
 d_{NID}(s,p)=\frac{max\{K(s|p^{*}), K(p|s^{*})\}}{max\{K(s), K(p)\}}=1-\frac{I(s:p)}{K(p)}.\label{eq:dnid}
\end{equation}
where in the last equality, we assumed, without loss of generality, that $K(p) > K(s)$. $d_{NID}$ complies, up to a constant value, with the triangle inequality, the symmetry axiom and the identity axiom. 

NID is not a Hamming type distance, characterizing only how much two sequence differ site by site instead, is an information based distance that quantifies how correlated are two sequence from the algorithmic perspective.  If two low-Kolmogorov random sequence can be, to a large extent, derived one from the other by a small sized algorithm, then the corresponding NID is small. If, on the other hand, they are not algorithmically correlated, then the NID is near one. If two sequence are highly random but are very similar site by site, they also will have a small NID value. 

As Kolmogorov randomness is uncomputable, a practical alternative to equation (\ref{eq:dnid}) is needed. Considering that $K(s|p)\cong K(s,p)-K(p)$, then $K(s|p)$ is roughly equal to $K(sp)-K(p)$ (there is a logarithmic correction term which is not significant for long enough sequences), $sp$ denotes the concatenation of both strings.

The NID distance can then be estimated from 
\begin{equation}
 d_{NID}(s,p)=\frac{K(sp)-min \{ K(s), K(p) \} }{max\{ K(s), K(p)\}}.\label{eq:dnidk}
\end{equation}
If the Kolmogorov randomness is estimated by using a compressing algorithm, then the normalized compressed distance (NCD) is now defined by
\begin{equation}
 d_{NCD}(s,p)=\frac{C(sp)-min \{ C(s), C(p) \} }{max\{ C(s), C(p)\}}\label{eq:dncd}
\end{equation}
 where $C(x)$ is the compressed size of the string $x$. Li et al \cite{li04}, as well as Emmert-Streib \cite{emmert10} used gzip and bzip2, as the compressing software, with no significant difference between the different compression softwares. 

Instead, if $s$ and $p$ have the same length, then we can cast equation (\ref{eq:dnidk}) in terms of the entropy density
\begin{equation}
 d_{NID}(s,p)=\frac{h(sp)-min \{ h(s), h(p) \} }{max\{ h(s),h(p)\}}.\label{eq:dnidh}
\end{equation}
 instead of a compressor, the normalized LZ76 complexity is now used to estimate the entropy density,
\begin{equation}
 d_{LZ}(s,p)=\frac{c_{LZ}(sp)-min\{c_{LZ}(s), c_{LZ}(p)\}}{max\{c_{LZ}(s), c_{LZ}(p)\}}.\label{eq:dlz}
\end{equation}
In what follows the $LZ$ will be dropped from $d_{LZ}$ every time it clutters notation, no ambiguity will arise. $d_{LZ}$ will have the same interpretation than $d_{NID}$  in as much as $c_{LZ}$ estimates the entropy density.

\section{Results}\label{sec:results}

Our initial configurations are random $10^4$ sites \footnote{The random configurations were taken from random.org site which draws random number from atmospheric noise from different geographic locations.} with entropy density near $h=1$ bits/site. For each configuration the system is let to evolve $10^4$ time steps.

\subsection{Dynamical evolution of ECA}\label{sec:dynevolveECA}

\begin{figure}[!t]
\centering
\includegraphics*[scale=2.5]{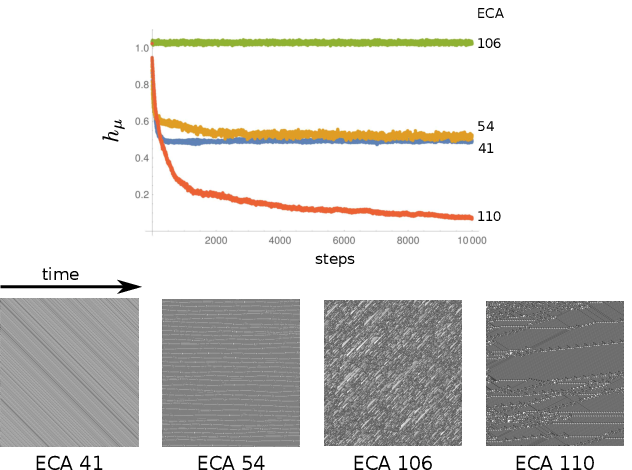}
\caption{Above: Entropy density evolution with time steps for several rules. Below: Sample runs of the plotted ECA rules. Time steps go from right to left son they can be compared with the entropy density evolution.
}\label{fig:classw4hmu}
\end{figure}

The entropy density evolution with time for all ECA was followed. Figure \ref{fig:classw4hmu} shows some examples. A group of rules evolves to low values of entropy density, this is certainly the case for all W1 ECA, but also for rules belonging to the other classes (See below). An interesting case is rule 110, which has been the focus of interest due to its universal computing capabilities \cite{wolfram02}. Rule 110 evolves to a small $h_{\mu}$ ($\approx 0.068$) value, with a long decaying time. Other rules settle into intermediate values of entropy density, such is the case for rule 41. Finally, others (e.g. rule 106) are characterized by almost no decrease of the entropy density with a, more or less, constant behavior of $h_{\mu}$ ( For the complete set of numerical results the reader is refered to table \ref{tbl:WECAtbl} in the Appendix).

Next, how ECA rules transfer information from the initial configuration as they evolve was studied. The $d_{LZ}$ between two consecutive configurations $s^t$ and $s^{t+1}$ was calculated along the time evolution, and the average was then obtained (denoted as $d^{p}_{LZ}$) and plotted parametrically against the final entropy density (figure  \ref{fig:prev_hmu}). In taking the average, the first $2\times 10^3$ steps were discarded to avoid the influence from the transient region. Clustering was performed using a hierarchical clustering algorithm with the Manhattan distance ($\sum |u-v|$) and a silhouette index as significant test\footnote{The significant test subdivides the data into successively more clusters looking for the first minimum of the silhouette statistic. The gap significant test was also used with no change in the clustering result. The gap test compares the dispersion of clusters derived from the data to that generated from a sample of null hypothesis sets. The actual implementation was done by using Wolfram Mathematica(c) FindCluster internal function.}. In this way two clusters were identified, one with the $d_{LZ}$ values around $1$, labeled dp3; and the  rest of the points in a second cluster (dp1+dp2). We further subdivided the last cluster in two, one comprising the W2 rules (dp2) and the other, made of W1 rules (dp1). The group of rules dp1 had zero $d^{p}_{LZ}$ value, signaling a complete transfer of information as the configurations evolves. These rules also show entropy densities near zero . The group of rules dp2, set to rules belonging to the WII class, are spread over the whole range of entropy density values, but show $d^{p}$ distance between $10^{-3}$ and $10^{-1}$. These rules follow a trend of decreasing $d^{p}$ with increasing entropy density. Finally, the third group of rules dp3, identified by the clustering algorithm, where all W3 and W4 and also some W2 rules are included, all show values of $d^{p}$ around one.  These rules loose, on the average, all information from one time step to the next. The W3 rules all show more or less the same average information distance value around $0.85$, and are grouped in three clusters for three different entropy density values. The W4 rules are also grouped in three clusters according to the entropy density value and has small dispersion of distance values. The W2 rules in the dp3 group show the largest dispersion of $d^{p}$ values spanning the whole range of $h_{\mu}$.

\subsection{Sensibility to initial configuration}\label{sec:initialconfECA}

Variations of the initial configurations were considered, all with a single site difference from the original random string. $3\times 10^2$ perturbed initial configurations were generated for each rule studied. The information distance between a perturbed configuration and the original initial configuration for each time step was computed. Several different behaviors were observed depending on the rule. 

\begin{figure}[!t]
\centering
\includegraphics*[scale=0.5]{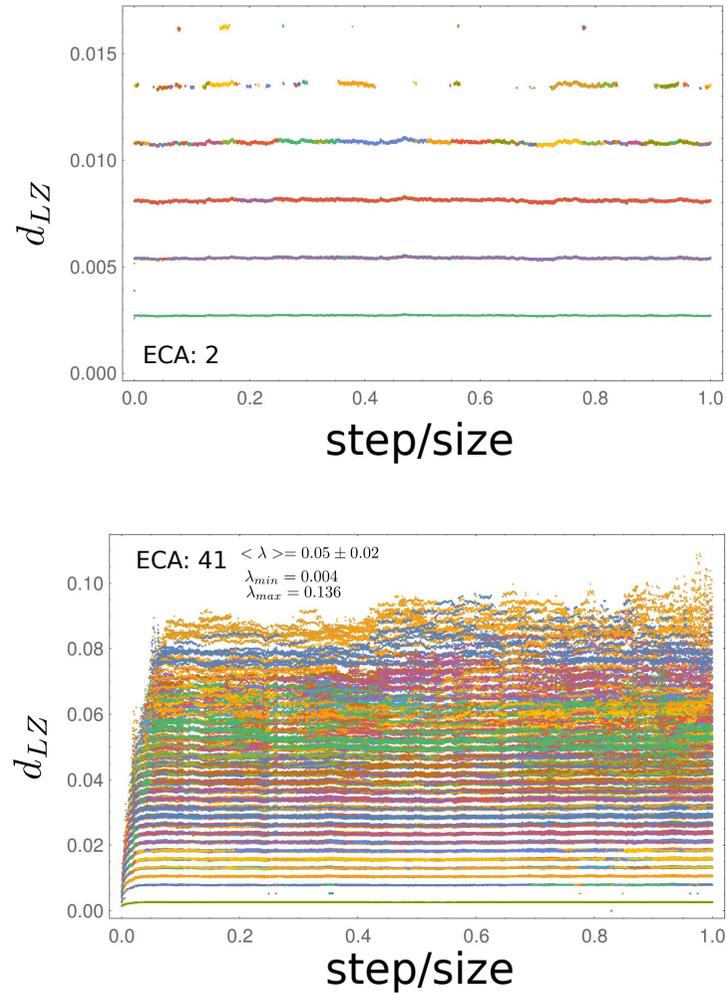}
\caption{The evolution of $d_{LZ}$ as a function of time steps (normalized by the total length of the sequence), for two rules showing almost no sensitivity to a single site perturbation of the initial configuration. For each rule $3\times 10^{2}$ perturbed initial configurations were chosen. For rule 41, $<\lambda>$ corresponds to the average value of the slope in the increasing region of $d_{LZ}$, while $\lambda_{min}$ and $\lambda_{max}$, are the smaller and larger slope found among all cases.
}\label{fig:lowncd}
\end{figure}

Some rules (1-13, 15, 19, 23-29, 33-38, 41, 42, 44, 46, 50, 51, 56, 58, 62, 72, 73, 74, 76-78, 94, 104, 108, 130, 132, 134, 138, 140, 152, 154, 156, 162, 164, 170, 172, 178, 200, 204, 232 and all equivalents)   exhibit almost no sensibility to the perturbation of the random initial configuration. The system dynamics quickly evolves towards a constant NCD with values below $10^{-1}$. In some cases this  jump is almost immediate (Figure \ref{fig:lowncd}a), while in others there is a small transient below 0.1 steps/size (Figure \ref{fig:lowncd}b)\footnote{Steps where normalized by the total spatial length of the sequences in order to show the abscissa of the plot independent the system size, actual steps follows by multiplying by the sequence length.}. 

\begin{figure}[!t]
\centering
\includegraphics*[scale=0.7]{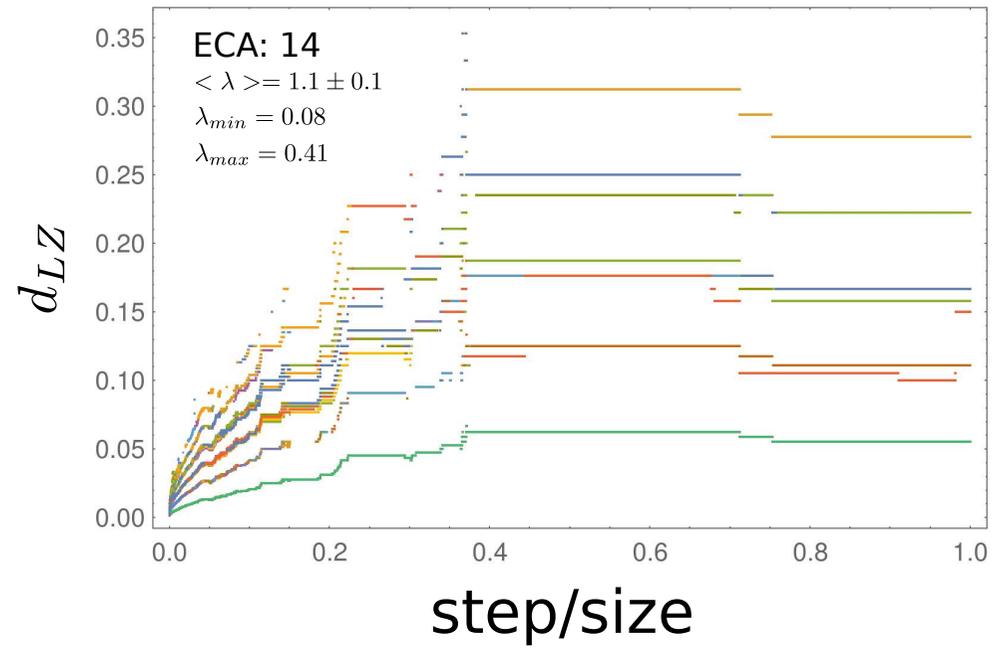}
\caption{Rule 14 shows an evolution of $d_{LZ}$ to intermediate values with a longer transient region. Conditions of the plot and notation follows figure \ref{fig:lowncd}.  
}\label{fig:rule14}
\end{figure}

Rule 14 (Figure \ref{fig:rule14}) shows a moderate sensibility to the perturbation of the random initial condition. In this case, there is a longer transient region, with a slope $\lambda$ between $[0.08, 0.41]$, a mean value of $<\lambda>=1.1$, the dispersion of slope values has a standard deviation of $0.1$. The $d_{LZ}$ values saturates around $0.35$. The set of nonequivalent rules exhibiting similar behavior are 25, 97, 121, 136, 184, 32, 40, 43, 142, 14, 128, 57  in increasing order of NCD value, all above $0.1$ and below $0.5$.

\begin{figure}[!t]
\centering
\includegraphics*[scale=0.7]{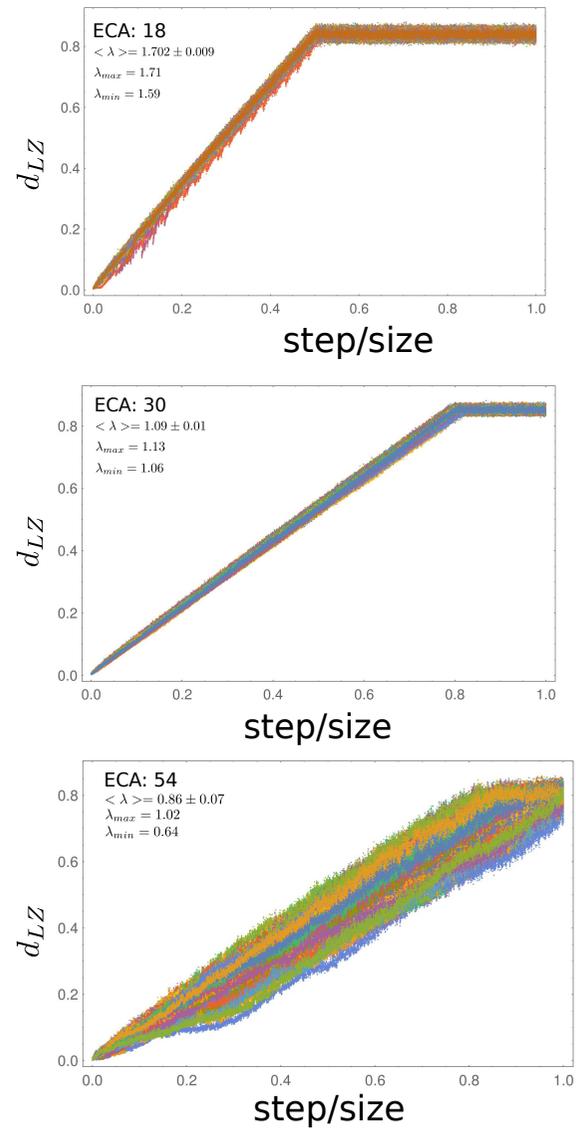}
\caption{The evolution of $d_{LZ}$ calculated for perturbed initial conditions, as a function of time steps for rules showing a linear behavior up to saturation. Rules 18 and 30 show less dispersion of the linear behavior among the perturbed cases. Conditions of the plot and notation follows figure \ref{fig:lowncd}. 
}\label{fig:largencd}
\end{figure}

A third group of rules exhibit a linear behavior of $d_{LZ}$ up to saturation (Figure \ref{fig:largencd}) above $0.5$.  In this group we find rules 18, 22, 30, 45, 54, 106, 110, 122, 126, 146, 160, 168 (and all equivalents). The slope of the linear law can vary, we found the largest slope for rule 18 with $<\lambda>=1.70$. They can also exhibit from a very small dispersion of slope values (Figure  \ref{fig:largencd}a,b), to a larger dispersion (Figure  \ref{fig:largencd}c).

\begin{figure}[!t]
\centering
\includegraphics*[scale=0.8]{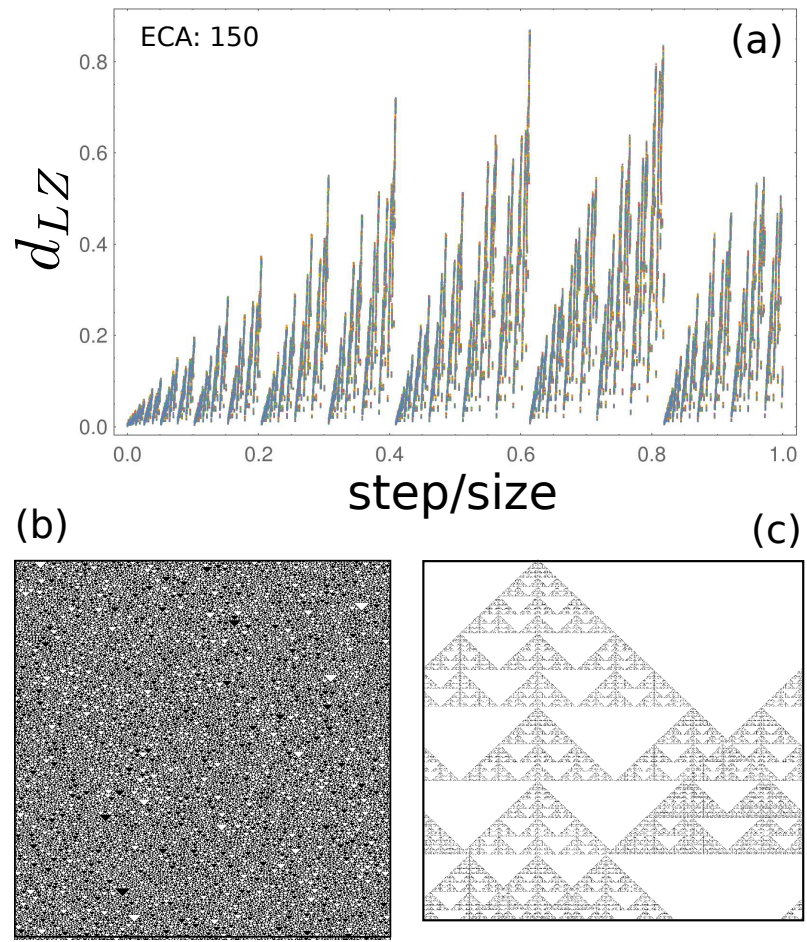}
\caption{Rule 150 fails to exhibit any monotonic behavior of  $d_{LZ}$ for perturbed initial conditions as a function of time steps. The plot is witnessing the fractal nature of the difference map showed in (c). The difference map was computed as the site-wise exclusive OR operation (XOR) of the spatiotemporal map of the rule. One of such spatiotemporal maps for the unperturbed initial configuration is shown in (b).
}\label{fig:rule150}
\end{figure}

Rules 60, 90, 105 and 150 (and all equivalents) exhibit a completely different behavior than the previous ones. Figure \ref{fig:rule150} shows the case for rule 150, the other rules in this group behave similarly. The $d_{LZ}$ value behaves somewhat cyclically. It starts increasing its value and then it falls again to near zero, to start increasing again. This graph of $d_{LZ}$ versus the time steps, has a self similar character. Rule 150 belongs to the W3 class. We plotted the difference map between the spatiotemporal evolution starting from the original random initial configuration and a perturbed initial configuration and is shown in figure  \ref{fig:rule150}c. The fractal nature of the difference curve is clear. The collapse of the  $d_{LZ}$ curve to almost zero value can be seen as those rows in the difference map with the apex of the triangle regions. In this ECA rules, no slope can be defined. 

\begin{figure}[!t]
\centering
\includegraphics*[scale=1.2]{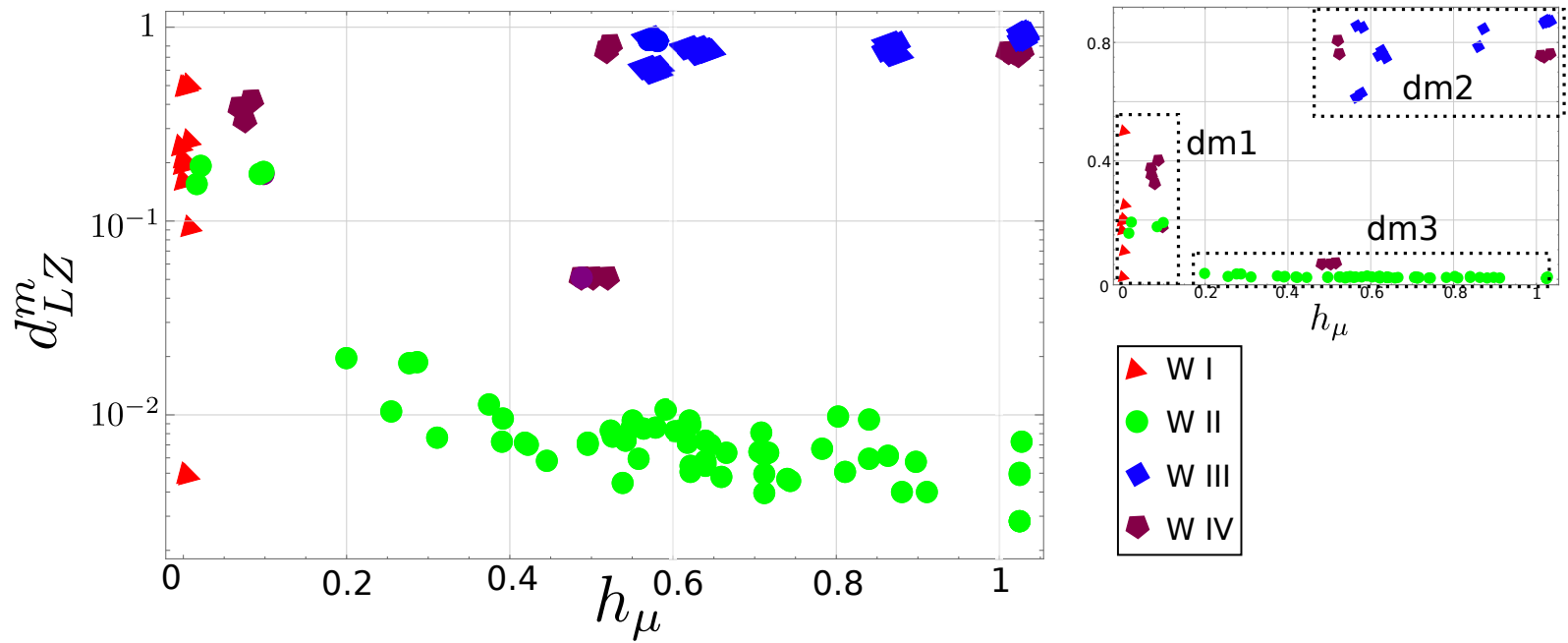}
\caption{The NCD as a function of entropy density as measured through $d^{m}_{LZ}$. $d^{m}_{LZ}$  is the average value of the maximum NCD for each perturbed initial configuration. The first two temporal configurations were discarded for each perturbed case.  The upper right plot is linear in both axis where three groups of rules can be identified. See text for further details.
}\label{fig:max_hmu}
\end{figure}

Returning to the first three groups of rules, figure \ref{fig:max_hmu} shows the plot of the largest NCD value $d^{m}$ between the sequence evolution of the original and perturbed initial condition (the first 2 time steps were discarded for this calculation) against the entropy density. Rules with low entropy density expand a region of $d^{m}$ from low values to intermediates one(dm1). Chaotic and complex pattern rules, have in general, high sensitivity to initial conditions (dm2). There are a third group of rules which expands a large interval of entropy density values and exhibit low maximum NCD (dm3). 
 
\begin{figure}[!t]
\centering
\includegraphics*[scale=1]{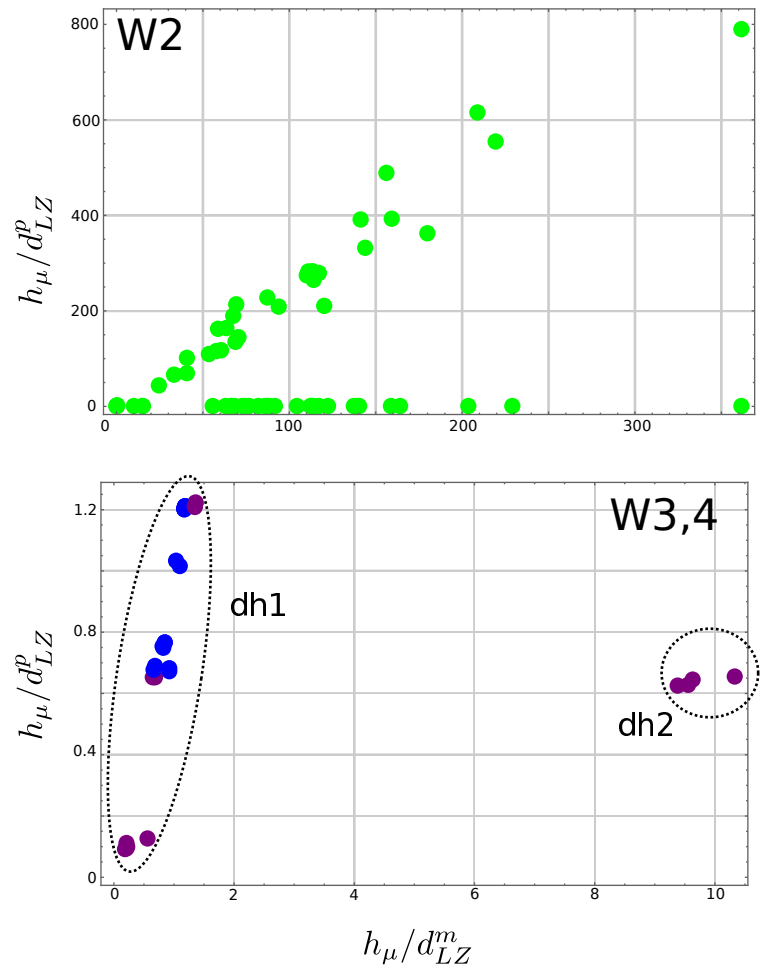}
\caption{Entropy density divided by $d^{p}$ plotted against $h_{\mu}/d^{m}$. For w2 class rules two regions are seen, one with almost constant behavior and, a second one showing a linear trend. For w3, 4 classes rule, a linear behavior is also identified, and second group consisting of a small cluster around some fixed values of the abscissa and the ordinate. 
}\label{fig:pre_max}
\end{figure}

Finally, entropy density was divided by $d^{p}$ and $d^{m}$ and the plot of the former against the  later is shown in figure \ref{fig:pre_max} for classes W2, W3 and W4. For the W2 class rules, two linear behaviors are observed, one with slope 0.006 bits/size which renders an almost independent ordinate with respect to the abscissa value (rules: 1, 3, 5, 6, 7, 9, 11, 15, 19, 23, 25, 26, 27, 28, 29, 33, 35, 37, 38, 43, 50, 51, 57, 62, 73, 74, 94, 108, 134, 142, 154, 156, 178, 184 and all equivalent rules), while another with a linear behavior with slope of $2.41$ bits/size (rules: 2, 4, 10, 12, 13, 24, 34, 36, 42, 44, 46, 56, 58, 72, 76, 77, 78, 104, 132, 130, 138, 140, 152, 162,  164, 170, 172, 200, 204, 232 and all equivalent rules ). For the rules in class W3 and W4,  also two groups can be distinguished. One (dh1) with a linear behavior with slope 0.94 bits/size (rules: 14, 18, 22, 30, 45, 54, 106, 110, 122, 126, 146 , and all equivalents), and another grouping (dh2) corresponding to rule 41 (and the equivalents: 97, 107 and 121).

\section{Discussion and conclusions}\label{sec:discuss}

\begin{figure}[!t]
\centering
\includegraphics*[scale=0.7]{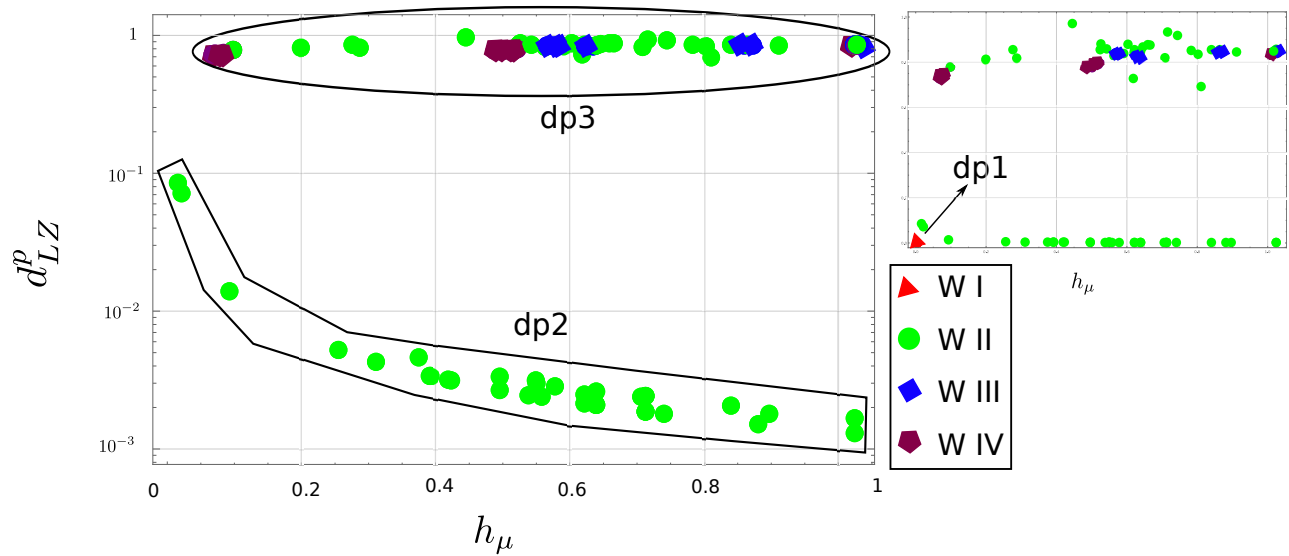}
\caption{NCD between time consecutive configurations as measured by $d^{p}_{LZ}$ plotted against the final entropy density. For each ECA rule, the first $2000$ time steps were discarded. The CA rules can be grouped in three classes. The main figure has a semi-log scale, while the smaller plot at the upper right corner is linear in both axis, showing that actually the rules belonging to the dp2 group have a NCD near zero, while the rules within the dp3 group show a larger dispersion that can be guessed from looking at the semi-log plot. In the semi-log plot, WI rules are not shown as their logarithmic diverges to infinity}\label{fig:prev_hmu}
\end{figure}

CA can be considered as computational systems performing some kind of computation over an initial (input) configuration. If viewed from this perspective, reduction of entropy density from the initial random sequence is associated with the capability of the system to structure the incoming information in an irreversible way. The system starts with a completely random input and randomness is lost (or not) during temporal evolution. Entropy density could then be considered a measure of the incapability of the system to recover the initial configuration from a later stage configuration.  Landauer principle \cite{landauer61} associates a minimal amount of energy dissipation to a single irreversible process (erasure) given by $k_{B}T\log 2$, where $k_{B}T$ is the Boltzmann factor. Zurek \cite{zurek89} showed that there is no energy cost of generating a string $s$ from its minimal program $s^{*}$ and viceversa. $h_{\mu}$ measures that portion (per bit) of the original string $s$ that can not be described by a shorter algorithm. Any reduction in $h_{\mu}$ must then be associated with irreversible computation and therefore Landauer principle applies. Those CA rules which result in a lower entropy density dissipate more energy of the system with respect to the initial condition. The asymptotic entropy density is then an (incomplete) measure of how energetically favorable is a given CA rule process, and of the stability of the final sequences with respect to other configurations starting from the same initial condition. Not surprisingly, class w1 rules exhibit the largest reduction of entropy density, yet Figure \ref{fig:prev_hmu}, as well as table \ref{tbl:WECAtbl}, show that rules with different behavior can result in a drastic lowering of the initial configuration entropy density: rules from Wolfram W1, W2 and W4 classes can result in fairly stable final configurations. Rules in class W1 and W2 evolve to more or less ordered configurations and the reduction of entropy comes as no surprise, but rules in class w4 evolve to highly complex configurations, that yet, in some cases, can exhibit low entropy density. One could discuss if this can be taken as a CA analogue to phase transformation from the liquid to the solid state. Class w1 and w2 automata reproduces more or less ideal crystallization but class w4 rules, with low final entropy rate, could be taken to simulate solidification process that lead to complex, aperiodic and disordered ground states. The fact that class w2 and w4 rules expand the full range of entropy density could indicate that this rules could be used to simulate different process leading to metastable phases. 

If we now consider the average NCD $d^{p}$ between two consecutive sequences in a CA driven configuration evolution, two interpretations can be given. The value of $d^{p}$ could be related to the 'speed' evolution of the configurations, were in one time step, for small $d^{p}$ value, few sites are changed. Another interpretation is related to how much the short time (immediate)  evolution seems in average like the consequence of an ``algorithmic agent''. Consider a configuration with low Kolmogorov randomness, then a small value of $d^{p}$ is related to the existence of a small algorithm that can describe both sequence. CA rules in group dp1 (figure \ref{fig:prev_hmu}) seems to be this case. Rules of class W2 included in group dp2 have intermediate values of $d^{p}$. Observe that the higher $h_{\mu}$ CA rules within this group have the lowest $d^{p}$ values and will therefore correspond to slow evolving sequences. Indeed, the identity rule 204 belongs to dp2 and, as expected, preserves the entropy density of the initial configuration and has the smallest value of $d^{p}$ within this group. Other rules in this corner of the dp2 group is rule 170 which was discussed by Emmert-Strieb \cite{emmert10} to be closer to rule 204 a fact confirmed in our analysis. In group dp3 are those rules where algorithmic correlation is lost from one time step to the next. Observe that this behavior can occur for the whole range of entropy density and the complete W3 and W4 classes are included here. In this group, few sites remain unchanged from one step to the next, and such changes can happen between highly disordered ($h_{\mu}\sim 1$) configurations or between sequences with low entropy density but essentially algorithmically uncorrelated.
 
In the analysis of CA sensitivity to initial conditions we also defined three groups (Figure \ref{fig:max_hmu}). Group dm1 belongs to low entropy density and shows $d^{m}$ values from very small to intermediate values $(< 0.5)$. All class W1 rules are included here as expected, but also some class W2 and W4 rules are also included. Rules in this group can be considered ``mild'' enough to initial conditions perturbation and therefore controllable in a useful calculation, and indeed the UTM rule 110 is in this group. Rules in group dm3 are insensitive to the initial conditions, and therefore unable to make  sense of a small change in the initial configuration, their behavior is very robust in a noisy environment. The bulk of class W2 rules are in this group as well as rule 41 (equivalent 97, 107 and 121) belonging to class W4. We should expect that their computational capabilities be limited by this insensitivity to the variations in the input sequence. Finally, group dp3 are those rules very sensible to a small perturbation in the initial sequence. The complete W3 class is included here, but also some class W4 rules. These rules are very perturbation sensitive and therefore should be unpractical for computation in a environment with even the smallest noise. Their evolution is not robust enough to be reliable.
 
Rules 60, 90, 105 and 150 all belonging to class W3 have a completely different behavior with respect to the sensitivity to initial conditions. In previous cases, a monotonic behavior could be found for $d^{m}$ as a function of time, but for these rules, this is impossible. No matter how long the system is left to evolve, the correlation between two different (perturbed) initial conditions will depend strongly on the specific time step and can be from nearly one to almost uncorrelated. All these rules do not reduce the amount of randomness of the initial input. The fractal nature of their difference evolution makes them an intriguing case with respect to the possibility of harvesting their computational capabilities.
 
It is not straightforward that there should be any relation between the information transfer in the time evolution of a CA and the sensibility to initial conditions perturbation. It is therefore interesting that, as figure \ref{fig:pre_max} shows the CA can be grouped according to certain relations between the magnitudes that characterize both behaviors. For the class W2 CA, we group two set of rules. In one, the $h_{\mu}/d^{p}$ quotient is almost constant, while for the second set a trend to a  linear behavior can be found.  For W3 and W4 class rules, two clusters are again identified, a small grouping around fixed values of both parameters and a trend to a linear behavior grouping both W3 and W4 rules. An interpretation of the different groups in term of computation capabilities is still under investigation and will be published elsewhere.
 
In summary, we have shown the power of Lempel-Ziv based analysis of CA spatiotemporal evolution. The use of different quantities that can be computed from the configurations itself, allows the analysis of collective behavior of CA in the spirit already put forward by Emmert-Strieb \cite{emmert10}. In spite of all the work done in CA classification, a completely consistent partition of rules in classes remains elusive and as several analysis, including this one, could suggest, perhaps such classification is intrinsically flawed. CA rules seems to have a much richer behavior than those that can be covered by a few parameters. The analysis performed sheds lights into the computational capabilities of CA as information processing rules and can be used to select CA rules amiable for simulating different physical process such as those related to solidification. The tools developed for the analysis of the CA are easily extendible to the study of other one dimensional dynamical systems an area we are starting to look into.

\section{Appendix}\label{sec:WECA}

\begin{longtable}{ll|ll|ll|ll}
\caption{ECA equivalent rules (taken from \cite{wolframalpha}).}\label{tbl:WECAequiv}\\
\hline  \\
rule & equivalents & rule & equivalents & rule & equivalents & rule & equivalents   \\
\hline \\
0  & 255           & 26 & 82, 167, 181 & 56  & 98, 185, 227   & 132 & 222            \\
1  & 127           & 27 & 39, 53, 83   & 57  & 99             & 134 & 148, 158, 214  \\
2  & 16, 191, 247  & 28 & 70, 157, 199 & 58  & 114, 163, 177  & 136 & 192, 238, 252  \\
3  & 17, 63, 119   & 29 & 71           & 60  & 102, 153, 195  & 138 & 174, 208, 224  \\
4  & 223           & 30 & 86, 135, 149 & 62  & 118, 131, 145  & 140 & 196, 206, 220  \\
5  & 95            & 32 & 251          & 72  & 237            & 142 & 212            \\
6  & 20, 159, 215  & 33 & 123          & 73  & 109            & 146 & 182            \\
7  & 21, 31, 87    & 34 & 48, 187, 243 & 74  & 88, 173, 229   & 150 & -              \\
8  & 64, 239, 253  & 35 & 49, 59, 115  & 76  & 205            & 152 & 188, 194, 230  \\
9  & 65, 111, 125  & 36 & 219          & 77  & -              & 154 & 166, 180, 210  \\
10 & 80, 175, 245  & 37 & 91           & 78  & 92, 141, 197   & 156 & 198            \\
11 & 47, 81, 117   & 38 & 52, 155, 211 & 90  & 165            & 160 & 250            \\
12 & 68, 207, 221  & 40 & 96, 235, 249 & 94  & 133            & 162 & 176, 186, 242  \\
13 & 69, 79, 93    & 41 & 97, 107, 121 & 104 & 233            & 164 & 218            \\
14 & 84, 143, 213  & 42 & 112, 171, 241& 105 & -              & 168 & 224, 234, 248  \\
15 & 85            & 43 & 113          & 106 & 120, 169, 225  & 170 & 240            \\
18 & 183           & 44 & 100, 203, 217& 108 & 201            & 172 & 202, 216, 228  \\
19 & 55            & 45 & 75, 89, 101  & 110 & 124, 137, 193  & 178 & -              \\
22 & 151           & 46 & 116, 139, 209& 122 & 161            & 184 & 226            \\
23 & -             & 50 & 179          & 126 & 129            & 200 & 236            \\
24 & 66, 189, 231  & 51 & -            & 128 & 254            & 204 & -              \\
25 & 61, 67, 103   & 54 & 147          & 130 & 144, 190, 246  & 232 & -              \\
\hline
\end{longtable}

\begin{longtable}{ll}
\caption{ECA rules (one representative per equivalence class) classified according to Wolfram classification scheme (taken from \cite{wolframalpha}).}\label{tbl:WECAclaf}\\
Class & rule \\
\hline \\
W1  & 0, 8, 32, 40, 128, 136, 160, 168  \\
W2  & all others \\ 
W3  & 18, 22, 30, 45, 60, 90, 105, 122, 126, 146, 150  \\
W4  & 41, 54, 97, 106,  110  
\end{longtable}

\begin{longtable}{lllll|lllll}
\caption{Final entropy density ($h_{\mu}$), maximum information distance between perturbed initial conditions ($d^{m}$) and, mean information distance between two consecutive steps ($d^{p}$) for all ECA rules (one representative per equivalence class) classified according to Wolfram classification scheme. Table is sorted according to Wolfram classes}\label{tbl:WECAtbl}\\
rule & class & $h_{\mu}$ & $d^{m}$ & $d^{p}$ & rule & class & $h_{\mu}$ & $d^{m}$ & $d^{p}$\\
\hline \\
 168 & \text{W1} & 0.0040 & 0.50 & 0.0019  & 44 & \text{W2} & 0.56 & 0.0060 & 0.0027 \\
 160 & \text{W1} & 0.0040 & 0.20 & 0.00094 & 43 & \text{W2} & 0.093 & 0.18 & 0.030 \\
 136 & \text{W1} & 0.0040 & 0.20 & 0.00091 & 42 & \text{W2} & 0.90 & 0.0057 & 0.0018 \\
 128 & \text{W1} & 0.0040 & 0.092 & 0.00049 & 38 & \text{W2} & 0.78 & 0.0067 & 0.85 \\
 40  & \text{W1} & 0.0040 & 0.17 & 0.00094 & 37 & \text{W2} & 0.56 & 0.0085 & 0.83 \\
 32  & \text{W1} & 0.0040 & 0.25 & 0.00054 & 36 & \text{W2} & 0.31 & 0.0076 & 0.0045 \\
 8   & \text{W1} & 0.0040 & 0.0050 & 0.00020 & 35 & \text{W2} & 0.65 & 0.0070 & 0.87 \\ 
 232 & \text{W2} & 0.62 & 0.0054 & 0.0023 & 34 & \text{W2} & 0.71 & 0.0063 & 0.0025 \\
 204 & \text{W2} & 1.0 & 0.0028 & 0.0013 & 33 & \text{W2} & 0.67 & 0.0064 & 0.88 \\
 200 & \text{W2} & 0.74 & 0.0047 & 0.0019 & 29 & \text{W2} & 0.91 & 0.0040 & 0.85 \\
 184 & \text{W2} & 0.021 & 0.19 & 0.072 & 28 & \text{W2} & 0.53 & 0.0077 & 0.88 \\
 178 & \text{W2} & 0.64 & 0.0056 & 0.85 & 27 & \text{W2} & 0.86 & 0.0062 & 0.85 \\
 172 & \text{W2} & 0.50 & 0.0072 & 0.0037 & 26 & \text{W2} & 0.84 & 0.0095 & 0.86 \\
 170 & \text{W2} & 1.0 & 0.0049 & 0.0017 & 25 & \text{W2} & 0.20 & 0.020 & 0.81\\
 164 & \text{W2} & 0.39 & 0.0096 & 0.0039 & 24 & \text{W2} & 0.58 & 0.0086 & 0.0030 \\
 162 & \text{W2} & 0.71 & 0.0064 & 0.0026 &  23 & \text{W2} & 0.62 & 0.0051 & 0.85\\
 156 & \text{W2} & 0.54 & 0.0073 & 0.86 & 19 & \text{W2} & 0.66 & 0.0048 & 0.88\\
 154 & \text{W2} & 1.0 & 0.0073 & 0.85 & 15 & \text{W2} & 1.0 & 0.0050 & 0.85\\
 152 & \text{W2} & 0.55 & 0.0094 & 0.0034 & 13 & \text{W2} & 0.42 & 0.0072 & 0.0036\\
 142 & \text{W2} & 0.098 & 0.18 & 0.77 & 12 & \text{W2} & 0.71 & 0.0040 & 0.0020 \\
 140 & \text{W2} & 0.71 & 0.0050 & 0.0022 & 11 & \text{W2} & 0.60 & 0.0083 & 0.88 \\
 138 & \text{W2} & 0.84 & 0.0060 & 0.0022 & 10 & \text{W2} & 0.71 & 0.0063 & 0.0025\\
 134 & \text{W2} & 0.59 & 0.011 & 0.84 & 9 & \text{W2} & 0.29 & 0.019 & 0.82\\
 132 & \text{W2} & 0.64 & 0.0055 & 0.0023 & 7 & \text{W2} & 0.52 & 0.0083 & 0.86\\
 130 & \text{W2} & 0.55 & 0.0086 & 0.0034 & 6 & \text{W2} & 0.62 & 0.0093 & 0.84\\
 108 & \text{W2} & 0.81 & 0.0051 & 0.69 & 5 & \text{W2} & 0.74 & 0.0045 & 0.92\\
 104 & \text{W2} & 0.26 & 0.010 & 0.0057 & 4 & \text{W2} & 0.54 & 0.0045 & 0.0026\\
 94 & \text{W2} & 0.62 & 0.0072 & 0.73 & 3 & \text{W2} & 0.72 & 0.0064 & 0.93\\
 78 & \text{W2} & 0.42 & 0.0070 & 0.0036 &  2 & \text{W2} & 0.50 & 0.0070 & 0.0034\\
 77 & \text{W2} & 0.64 & 0.0058 & 0.0023 & 1 & \text{W2} & 0.45 & 0.0058 & 0.97\\
 76 & \text{W2} & 0.88 & 0.0040 & 0.0016 & 146 & \text{W3} & 0.58 & 0.85 & 0.84\\
 74 & \text{W2} & 0.71 & 0.0081 & 0.82 & 126 & \text{W3} & 0.64 & 0.75 & 0.83\\
 73 & \text{W2} & 0.80 & 0.0098 & 0.84 & 122 & \text{W3} & 0.63 & 0.77 & 0.83\\
 72 & \text{W2} & 0.39 & 0.0073 & 0.0036 & 45 & \text{W3} & 1.0 & 0.87 & 0.85\\
 62 & \text{W2} & 0.28 & 0.018 & 0.86 & 30 & \text{W3} & 1.0 & 0.87 & 0.85\\
 58 & \text{W2} & 0.37 & 0.011 & 0.0056 & 22 & \text{W3} & 0.86 & 0.79 & 0.85 \\
 57 & \text{W2} & 0.016 & 0.16 & 0.10 & 18 & \text{W3} & 0.58 & 0.63 & 0.84\\
 56 & \text{W2} & 0.62 & 0.0090 & 0.0029 & 110 & \text{W4} & 0.068 & 0.38 & 0.74\\
 51 & \text{W2} & 1.0 & 0.0028 & 0.86 & 106 & \text{W4} & 1.0 & 0.75 & 0.84\\
 50 & \text{W2} & 0.64 & 0.0057 & 0.85 & 54 & \text{W4} & 0.52 & 0.81 & 0.80\\
 46 & \text{W2} & 0.64 & 0.0073 & 0.0028 & 41 & \text{W4} & 0.49 & 0.051 & 0.78 \\
    &           &      &        &        & 14 & \text{W4} & 0.098 & 0.18 & 0.78
\end{longtable}

\section{Acknowledgments}

We would like to thank an anonymous referee for its suggestions that greatly improved the article.

This work was partially financed by FAPEMIG under the project BPV-00047-13. EER which to thank PVE/CAPES for financial support under the grant 1149-14-8. Infrastructure support was given under project FAPEMIG APQ-02256-12.


\pagebreak

\end{document}